\documentstyle[12pt,aps,psfig,multicol,epsf,amssymb]{revtex}
\begin{document}
\draft
\title{Ordering of dipolar Ising crystals}
\author{Julio F. Fern\'andez} 
\address{ICMA, CSIC and Universidad de Zaragoza, 50009-Zaragoza, Spain}
\author{Juan Jos\'e Alonso}
\address{Departamento de F\'{\i}sica Aplicada I, Universidad de M\'alaga\\ 
29071-M\'alaga, Spain\\}
\date{\today}
\maketitle

\begin{abstract}     
We study Ising systems of spins with dipolar interactions.
We find a simple approximate relation for the interaction energy between
pairs of parallel lattice columns of spins running along the Ising spin direction.
This relation provides insight into the relation between lattice geometry
and the nature of the ordered state. 
It can be used to calculate ground state energies.
We have also obtained ground state energies and ordering temperatures $T_0$
from Monte Carlo simulations. Simple empirical relations, that give
$T_0$ for simple and body centered tetragonal lattices in terms of lattice
parameters are also established. Finally, the nature of the ordered state
and  $T_0$ are determined for Fe$_8$ clusters, which crystallize
on a triclinic lattice.

\end{abstract}
 

Dipolar interactions can lead to order at low temperature in
magnetic \cite{reich} as well as in ferroelectric systems \cite{luo,vug}.
Interacting dipolar Ising systems (DIS) have recently become the subject of special
interest because spin quantum tunneling has been observed
in them \cite{MQT,sangregorio}. It can take place
at temperatures that are well below
their ordering temperature $T_0$ (see below). The type of order that ensues and $T_0$
are not trivially determined owing to the
long range nature of dipolar interactions and to their sign changes.
Luttinger and Tisza were able to show long ago that the type of ordering
depends on the geometry of the lattice \cite{lut}. The theory was
later generalized by Niemeijer and Bl\"otte \cite{nie}. It enables 
one to determine, through laborious calculations, the ground state
energies and types of order for simple
Bravais lattices with up to two identical magnetic dipoles per unit cell.
Ordering temperatures are known from Monte Carlo (MC) simulations 
for simple cubic lattices \cite{binder}.
Unfortunately, symmetry forbids the existence of {\it cubic}
dipolar {\it Ising} systems in nature. There are however magnetic systems,
such as the so called single-molecule
magnets\cite{lis} as well as some rare earth sulfides\cite{cooke},
which (1) crystallize in other lattice structures and have sizable single-axis
anisotropies, and (2) the organic material around these clusters precludes
all magnetic interactions except for the magnetic dipolar one.
Consequently, they are expected to behave as DIS, and they indeed seem to
do so.\cite{cooke,bell,MC} Magnetic relaxation of some of these
single-molecule magnet crystals (Mn$_{12}$ and Fe$_8$)
have recently been simulated in connection
with spin quantum tunneling experiments \cite{MC}, but their thermal
equilibrium properties have not, as far as we know, been studied.

The aim of this report is twofold. First, we wish to point out
that there is a simple approximate relation for the interaction energy between
pairs of parallel lattice columns of spins running along the Ising spin direction,
and that the nature of the ordered state follows from this relation.
Second, we report Monte Carlo results for the ordering temperatures
of DIS on cubic lattices,
simple and body centered tetragonal lattices (that include
Mn$_{12}$ cluster crystals \cite{lis}), as well as on the lattice (triclinic) on which
Fe$_8$ crystallizes \cite{agnew}.  

We treat here systems of spins, on lattices to be specified below,
that are restricted to point along the $z$-axis,
and interact among themselves only through magnetic dipole interactions.
More specifically, let $S_i=\pm S$ be a spin on the $i$-th site of a simple cubic
or tetragonal lattice, with Hamiltonian
\begin{equation}
{\cal H}=-{v_0\over 2}\sum_{ij}V_{ij}s_is_j,
\label{model}
\end{equation}
where $s_i=S_i/S$, $V_{ij}=(a_z/r_{ij})^3(1-3z_{ij}^2/r_{ij}^2)$,
$a_z$ is the nearest neighbor distance along the $z$-axis, $r_{ij}$ is
the distance between sites $i$ and $j$,
\begin{equation}   
v_0={\mu_0\over {4\pi}}(g\mu_B S)^2/a_z^3,
\label{varep}
\end{equation}
$g$ is the gyromagnetic ratio, $\mu_B$ is the Bohr magneton, and
$\mu_0=4\pi\times 10^{-7}$ in SI units.

>From here on, unless explicitely specified to be in Kelvin,
energies and temperatures are given in terms of $v_0$.
It may be helpfull to keep in mind that,
$(\mu_0/4\pi)\mu_B^2/(1\AA)^3\simeq 0.622$K.

We have used the standard Metropolis algorithm in our
Monte Carlo simulations \cite{met}. We have mostly
used periodic boundary conditions (PBC). For PBC,
a spin at site $i$ is allowed to interact only with spins that lie within
a system sized box centered on site $i$.
We have also simulated some systems with free boundaries, some shaped
like a box, and some bounded by spherical surfaces. 
With free boundaries, all spins in the system are allowed to interact. 
The results obtained are in agreement with the expected behavior that
follows from Griffith's theorem: that the thermodynamic limit is
independent of boundary conditions and of system shape if
no external field is applied \cite{grif}.

We next find the field $B_\lambda$ that an infinitely long column
of spins pointing up, as in Fig. 1, produces a distance $a_x$ away from it.
First note that, as can be easy checked,  $B_\lambda$ would vanish if
the moment density were {\it uniformly distributed} rather than on a lattice.
An exact expression,
as an expansion in powers of modified Bessel functions, can be obtained
for $B_\lambda$ by writing the magnetic dipole linear density
along a column as a Fourier series
with period and phase that depend on $a_z$ and $d_z$, respectively
(see Fig.1).
The zeroth order term (which corresponds to a constant magnetic
dipole density) does not contribute.
The leading term is of the form
$(a_x/a_z)^{-1/2}\exp (-2\pi a_x/a_z)\cos (2\pi d_z/a_z)$.
We have checked numerically that
\begin{equation}
B_\lambda\simeq -\left({0.1357\over{\sqrt{a_x/a_z}}}\,B_0\right)
\exp [2\pi(1-a_x/a_z)]\cos (2\pi d_z/a_z),
\label{lines}
\end{equation}
where 
\begin{equation}
B_0\equiv {\mu_0\over {4\pi}}{{g\mu_B S}\over{a_z^3}},
\end{equation}
is within $1\%$ of exactness for $a_x/a_z >0.8$.
(Deviations from exactness increase as $a_x/a_z$
decreases, up to $11\%$ for $a_x/a_z =0.5$.). A $B_\lambda <0$ value
means that a column of spins up produces a field that points down.

On the other hand, the field $B_s$ of an infinite column of spins,
all pointing up, at one of its
own sites is easily found to be $B_s=4.808...B_0$.
This is much larger than $B_\lambda$ for any reasonable
distance between columns, and gives therefore rise to ferromagnetic order
within each column of spins. As shown below, this picture holds for tetragonal
lattices for $a_x/a_z \gtrsim 0.6$, but may break down for smaller values
of $a_x/a_z$.

One therefore expects the ordered state on simple cubic
and primitive tetragonal lattices, for which $d_z=0$ in
Eq. (\ref{lines}), to be a two-dimensional {\bf anti}-ferromagnetic array
of ferromagnetically ordered columns spins
as shown in Fig. 1b. On the other hand body-centered cubic and tetragonal
lattices, in which $d_z=a_z/2$ for nearest neighbor columns, are expected to
order ferromagnetically. This is as originally predicted by Luttinger and
Tisza \cite{lut}. 

Making use of Eq. (\ref{lines}) and the value of $B_s$,
the ground state energy $\varepsilon_0=-(1/2)(g\mu_BS)(B_\lambda+B_s)$
can be easily calculated for a simple cubic or tetragonal lattice. It is, 
\begin{equation}
\varepsilon_0\simeq -2.404v_0-1.8\,g\mu_BS\,B_\lambda.
\label{e0}
\end{equation}
This result is shown in Fig. 2, with data points obtained from Monte Carlo
simulations, as a function of the basal plane lattice constant
$a_x$. For ferromagnets the calculation of $\varepsilon_0$ is somewhat more
involved and is not attempted here. This is because
for lines of {\it finite} length
$L_z$, equation (\ref{lines}) is applicable only if $\ln (L_z/a_z)\gg a_x/a_z$.
For $\ln (L_z/a_z)\lesssim a_x/a_z$, $B_\lambda<0$, independently of $d_z$,
which gives rise to magnetic domains in ferromagnets. Equation (\ref{lines})
can be used to obtain the ``wall energy'' that counterbalances
``magnetostatic'' energies in domain size and ground state energy calculations
\cite{kittel}.

Data points obtained from MC simulations for the
ordering temperature of DIS on primitive tetragonal lattices are also shown in
Fig. \ref{C2}a. The equation
\begin{equation}
T_0\simeq 2.5\, v_0 (a_z/a_x)^{1.7}
\label{fit}
\end{equation}
provides the fit shown in Fig. \ref{C2}a.

For body centered tetragonal lattices, ferromagnetic order ensues, as expected
from the fact that $d_z=a_z/2$ then. The ground state energy is not simply
obtained then, since there are long range contributions. Data points
exhibited in Fig. \ref{C2}b, show that the ordering
temperature follows a slightly different rule from Eq. (\ref{fit}).
As shown in the table, the best fit to $T_0$ is then
$T_0\simeq 5.8\, v_0 (a_z/a_x)^{2.0}$.

Thermal equilibrium results obtained by MC simulations of DIS
on simple cubic lattices of (1) up to $16\times 16\times 16$ spins
with PBC and (2) $2109$ and $4169$ spins on spherical systems
with free boundary conditions
are shown in Figs. \ref{C3}a, \ref{C3}b, and \ref{C3}c.
These results, together with results (not shown) for smaller systems,
as well as for box shaped systems with free boundary conditions of
different sizes, lead to ordering temperatures and ground state energies
that are, within
statistical errors, independent of shape and boundary conditions.
Similar results have been obtained for simple and body centered
tetragonal lattices, from which the data points
for $T_0$ shown in Figs. \ref{C2}a and \ref{C2}b, respectively,
were obtained.

Triclinic crystals of Fe$_8$ clusters provide an interesting example \cite{agnew}.
Their lattice geometry is exhibited in Fig. \ref{C4}. Note that
lattice sites on columns marked with a diamond, circle, square, etc.
are displaced along the $z$-axis a distance
$d_z\approx a_z/4$ with respect to sites on columns marked with a
circle, square, triangle, etc., respectively. Thus, while all columns on the same
row on the right hand side of Fig. \ref{C4} 
interact antiferromagnetically, two columns on adjacent rows do not interact,
and two rows on alternate rows (e.g., on rows marked with circles and triangles)
interact ferromagnetically. This implies that any given row of columns in
Fig. 4 orders antiferromagnetically, but there is near degeneracy (if only nearest
nearest neighbor interactions are taken into account) as to
how adjacent rows order with respect to each other. How rows order with respect
to each other is quite likely determined by longer range interactions, but  
this question is beyond the scope of this report.
To find out how the system actually orders, we turn to MC simulations.
The ordered state is depicted in Fig. 4 as follows: $\blacklozenge$,
$\bullet$, $\blacksquare$, $\blacktriangle$, and $\blacktriangledown$ stand
for columns with spin up, while $\lozenge$, $\circ$, $\square$, $\bigtriangleup$,
and $\bigtriangledown$ stand for columns with spins down.
This system orders at a temperature given in table I.

It is worth remarking that Eq. (\ref{lines}) is expected to
be applicable whether dipoles are point like or extended,
or whether they are magnetic or electric, except that the
constant $0.1357B_0$ is different in each case.
It's validity only requires that the wave number
of the leading fourier component of
the dipolar density along the relevant columns be equal to the smallest
reciprocal lattice vector for the colunms. Furthermore, Eq. (\ref{lines})
is applicable to other crystal structures, such as hexagonal, not explicitely
treated here. It follows, for instance, that
DIS, whether magnetic or electric, order
antiferromagnetically in a primitive hexagonal
structure, but order ferromagnetically in a hexagonal closed packed structure.

In summary, we have shown that Eqs. (\ref{lines}) and (\ref{e0})
provide simple approximate relations for the interaction energy between
pairs of parallel lattice columns of spins running along the Ising spin direction,
and that the nature of the ordered state follows from this relation.
>From this simple relation, we have obtained
ground state energies.
We have also obtained ground state energies and ordering temperatures
from Monte Carlo simulations. A simple empirical relation, Eq. (\ref{fit}) gives
results for the ordering temperatures
of DIS on simple and body centered tetragonal lattices.
These results, as well as results for Fe$_8$ clusters that crystallize on
a triclinic lattice, are summarized on table I. (Substituting $g\simeq 2.0$,
$S=10$, and lattice constants (see Fig. \ref{C4}) of Fe$_8$
cluster crystals, gives $T_0\simeq 170$ mK. Similarly, $T_0\simeq 430$ mK is
obtained for Mn$_{12}$ acetate, which crystallizes in a primitive tetragonal
lattice with $a_z=12.39\AA$ and $a_x=17.32\AA$ \cite{lis}.)

\acknowledgments

We are grateful to Dr. F. Luis for many enlightening comments,
to Prof. N. Dalal for valuable information about the relation
between the spin Hamiltonian axes and crystalline axes,
and to Prof. P. Rikvold and Prof. M. Novotny for their hospitality
during a one month visit at the Supercomputing Computations Research
Institute at Florida State University by one of us (J. F. F.).
J. F. F. and J. J. A. are also grateful
to DGESIC of Spain for grants No. PB99-0541 and PB97-1080, respectively.

\newpage
\begin{table}
\caption{Type of ordered state, ordering temperature $T_0$, and ground state
energy $\varepsilon_0$, for dipolar
Ising systems on various lattices. All energies and temperatures are
given in terms of $v_0$, where
$v_0=(\mu_0/4\pi )(g\mu_B S)^2/a_z^3$, and $(\mu_0/4\pi)\mu_B^2/(1\AA)^3\simeq 0.622$K.}
\begin{tabular}{l c c c l} 
LATTICE                  & ORDER      &$T_0$  &$\varepsilon_0$ & validity\\ \tableline
sc$^{1,2}$ & AF$^3$         & 2.50(5)   &   -2.68(1) &       \\
BCC$^1$ & F                & 5.8(2)  &-4.0(1)    &      \\
FCC$^1$            & F                    & 11.3(3) & -7.5(1)  &         \\
Primitive tetragonal  & AF$^3$     & 2.5$(a_z/a_x)^{1.7}$ & Eqs. (3-5) & $a_x/a_z\gtrsim 0.6$   \\
BC tetragonal$^5$  & F                & 5.8$(a_z/a_x)^{2.0}$  &Fig. 2b     & $a_x/a_z \gtrsim 0.6$        \\
Fe$_8$ (triclinic)           & AF$^4$                    & 1.9(1) & -2.73(3) &             \\ \tableline
\multicolumn{4}{l}{1. Type of order and $\varepsilon_0$ 
(but not $T_0$) first given in Ref. \cite{lut}.}\\
\multicolumn{4}{l}{2. $T_0$, from MC simulations, given in Ref. \cite{binder}.}\\
\multicolumn{4}{l}{3. AF order depicted in Fig. 1.}\\
\multicolumn{4}{l}{4. AF order depicted in Fig. 4.}\\
\multicolumn{4}{l}{5. $T_0$, from MC simulations, given in Ref. \cite{xu} for $a_x/a_z=0.5$.}\\
\end{tabular}\label{tabla}\end{table}

\newpage
\begin{figure}
\centerline{\psfig{figure=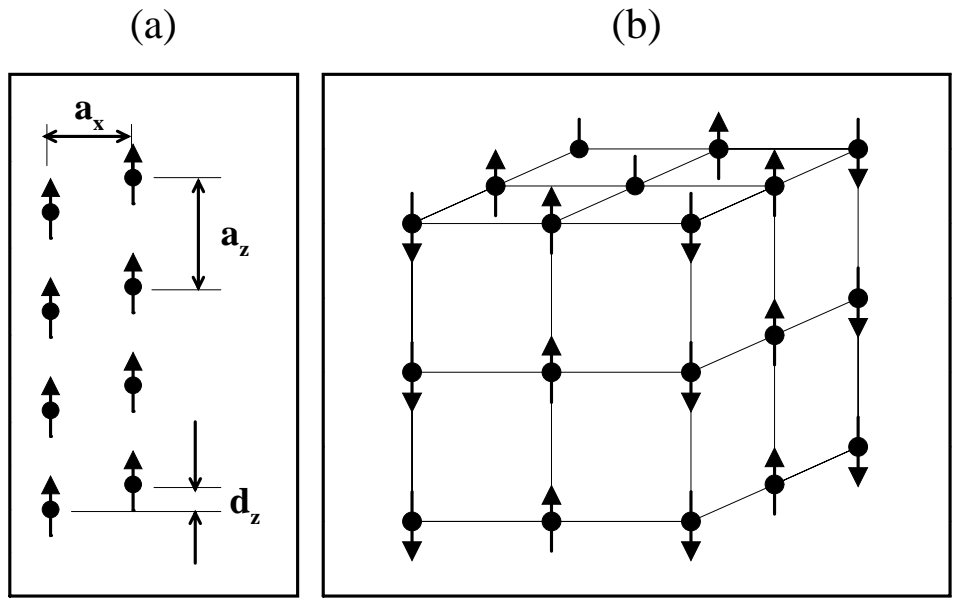,width=14cm,angle=0}}
\caption{(a) Parallel lines of spins,
showing distances $a_z$, $a_x$, and $d_z$.
(b) Antiferromagnetic order in simple cubic lattice.}
\label{C1}
\end{figure}
\newpage
\begin{figure}
\centerline{\psfig{figure=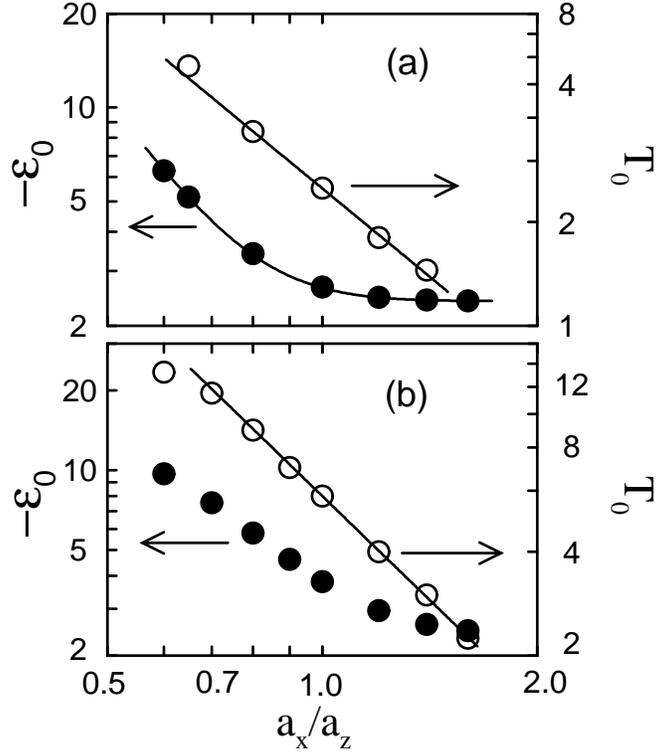,width=10cm,angle=0}}
\caption{(a) Semilog plot of the thermal equilibrium
ordering temperature $T_0$ and of the ground state energy $\varepsilon_0$ versus $a_x/a_z$
of the magnetic dipole model on a
tetragonal lattice, where $a_x$ and $a_z$ are the nearest
neighbor distances along the basal plane and $c$-axis respectively.
$\circ$ and $\bullet$ stand for temperature and energy, respectively.
The continuous line for energy follows from Eq. (\ref{e0}). The straight line
through the temperature data points is a fit; the equation for it is (\ref{fit}).
The results for $a_x/a_z = 1.4$
are obtained from columns of $32\times 32$ spins
on the basal plane by $128$ spins along the $c$-axis; the result for
$a_x/a_y=1.5$ is from a column of $8\times 8\times 512$ spins.
All values of $\varepsilon$ and $T$
are given in terms of $v_0$, defined in Eq. \ref{varep}. 
(b) Same as in part ``a'' but for
a dipole model on a BC tetragonal lattice. The straight line
through the temperature data points is a fit; the equation 
for it is 5.8$(a_z/a_x)^{2.0}$.}
\label{C2}
\end{figure}
\newpage
\begin{figure}
\centerline{\psfig{figure=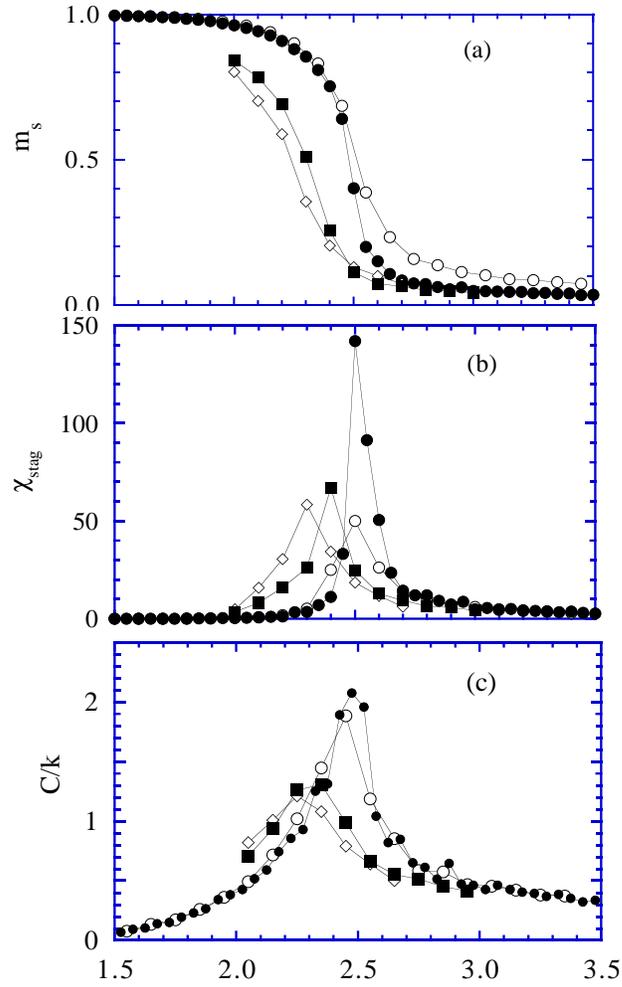,width=10cm,angle=0}}
\caption{(a) The thermal equilibrium staggered magnetization $m_s$
versus temperature.
$\circ$ and $\bullet$ stand for systems of $8\times 8\times 16$ and
$16\times 16\times 16$ spins, respectively, on a SC lattice with PBC. 
$\diamond$ and $\blacksquare$ stands
for spherically shaped systems of $2109$ and $4169$ spins, respectively, with free end
boundary conditions.
(b) Same as in part ``a'' but for the staggered susceptibility $\chi_{s}$.
(c) Same as in parts ``a'' and ``b'', but for the specific heat.}
\label{C3}
\end{figure}

\newpage
\begin{figure}
\centerline{\psfig{figure=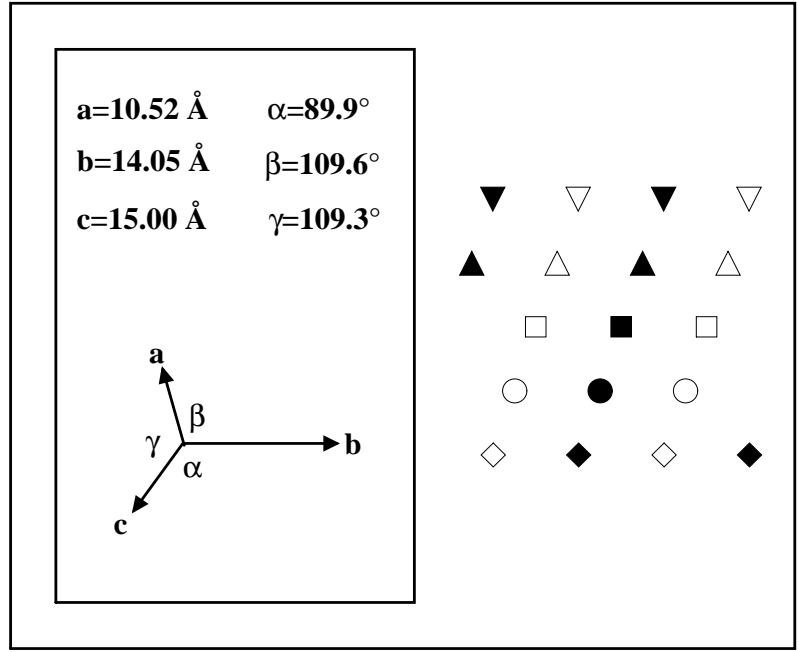,width=14cm,angle=0}}
\caption{Lattice constants of Fe$_8$ cluster crystals are
shown on the left hand side. The lattice
parameters are taken from Ref. [13]. On the right hand side, the lattice
as seen looking down the $b$-axis. Anisotropy constrains all spins
to be either up or down along the $b$-axis for $T< 1$K 
in this system. The
$b$-axis is therefore our $z$-axis.}
\label{C4}
\end{figure}

\end{document}